\documentclass[aps,twocolumn,pra,superscriptaddress,showpacs,tightenlines]{revtex4}
\usepackage{amssymb}
\usepackage{amsmath}
\usepackage{graphicx}
\usepackage{epsfig}
\usepackage{subfigure}
\usepackage{amsfonts}

\begin{document}

\title{Coherently manipulating flying qubits in a quantum wire with a
magnetic impurity}
\author{Lan Zhou}
\affiliation{Key Laboratory of Low-Dimensional Quantum Structures and Quantum Control of
Ministry of Education, and Department of Physics, Hunan Normal University,
Changsha 410081, China}
\author{Jing Lu}
\affiliation{Key Laboratory of Low-Dimensional Quantum Structures and Quantum Control of
Ministry of Education, and Department of Physics, Hunan Normal University,
Changsha 410081, China}

\begin{abstract}
We study the effect of a magnetic impurity with spin-half on a single
propagating electron in a one-dimensional model system via the tight-binding
approach. Due to the spin-dependent interaction, the scattering channel for
the flying qubit is split, and its transmission spectrum is obtained. It is
found that, the spin orientation of the impurity plays the role as a spin
state filter for a flying qubit.
\end{abstract}

\maketitle



\section{\label{Sec:1}Introduction}

Qubits are the quantum state of given physical systems. There are two kinds
of qubits, the first kind is called stationary qubits which are fixed into
space, the second kind is the so-called flying qubits whose position changes
in time. To faithfully transfer quantum information, individual qubit
control would be desirable. Controlling qubits are implemented by a sequence
of logical operations or quantum gates. For stationary qubits, quantum gates
are performed by choosing a proper time, however, a logic operation has to
be fixed into space for flying qubits.

Nowadays, more attention has been paid on the use of the flying qubit due to
the advantage that it allows one to entangle distant stationary qubits that
never interacted directly \cite{KwekL95,KwekA77}. Therefore, there are with
keen interest in seeking new device for controlling flying qubits \cite%
{NJcavity,fanpaper,Sun1,EITqssp,Lukin-np,gongzr,Kimble,ZGLSN,zhoucavity,YangA73,flyJP18,flyEL74,flyB77}
. Photon is the archetypal flying qubit, which is special ideal for long
distance communication due to its high speed, strong stability and minor
loss. Therefore quantum devices that enable new filtering and switching
functions have been proposed, such as single photon transistor~\cite%
{Lukin-np,gongzr}, quantum switch~\cite{fanpaper,Sun1,ZGLSN,EITqssp}. The
spin of a propagating electron is an alternative leading candidate for a
flying qubit~\cite{YangA73,flyJP18,flyEL74,flyB77} in short distance
communication mostly due to the intrinsic nature of electron spin, namely
that the spin degree of freedom is well isolated from the environment. The
ability to manipulate an mobile spin qubits is indispensable for the
potential application in quantum information processing. A singlet spin
filter has been proposed by the Coulomb interaction between a flying qubit
and the trapped electron in a weak confining potential \cite{flyEL74}. Later
on it was generalized to a potential well with multiple bound orbitals \cite%
{flyB77}. In this paper, we propose a scheme whereby a single propagating
electron is subject to a spin-spin interaction with a magnetic impurity in a
quantum wire. The local external magnetic field removes the degeneracy. Then
the spin-dependent scattering behavior of the propagating electron has been
investigated via the tight-binding approach. Due to the spin-flip
interaction between a single propagating electron and the magnetic impurity,
a total reflection of the propagating electron has been found by a proper
choice of the injection energy of the electron, i.e. a spin filter is formed
which filters the electron spin antiparallel to that of the magnetic
impurity.

This paper is organized as follow. In Sec.~\ref{Sec:2}, we introduce our
system, a flying qubit in one-dimensional (1D) model systems with a magnetic
impurity inserted locally. In Sec.~\ref{Sec:3}, the spin-induced
multichannels is found for this 1D quantum wire. And via single-electron
configurations, the quantum transport of the flying qubit has been analyzed.
A spin switching mechanism is provided on the Fano resonance, which is cause
by the coupling between the bound state outside one energy band and the
continuum in the other band. Conclusions are summarized at the end of the
paper.

\section{\label{Sec:2}a flying qubit with a magnetic impurity}

\begin{figure}[tbp]
\includegraphics[bb=86 629 515 698, width=7 cm,clip]{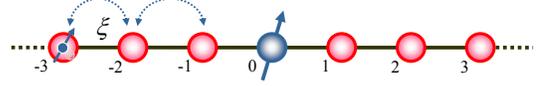}
\caption{(Color online). Schematic of the perfect quantum wire in one
dimension, where a magnetic impurity with spin-$\frac{1}{2}$ is embedded in
the 0th site.}
\label{scatfly:1}
\end{figure}

The model we considered in this paper is shown in Fig.~\ref{scatfly:1}.
Thanks to the success of nanofabrication techniques in producing extremely
small quantum objects, this kind of quantum wires are now close to reality
either by small semiconductor and metal structures as a fabricated quantum
dot array \cite{GSMMK98,GGK1998}, or by the scanning tunnel microscope~\cite%
{STM91,STMrev} as the atomic chain.

The system in Fig.~\ref{scatfly:1} is a one-dimensional wire, which also has
a impurity embedded in a host lattice. The Hamiltonian of this system
consists three parts
\begin{equation}
H_{c}=\sum_{\sigma }H_{qw}^{c\sigma }+H_{fr}^{c}+H_{ex}\text{,}  \label{mI-1}
\end{equation}%
where
\begin{eqnarray}
H_{qw}^{c\sigma } &=&\sum_{j}\omega a_{j\sigma }^{\dag }a_{j\sigma
}+\sum_{j}\xi \left( a_{j\sigma }^{\dag }a_{j+1\sigma }+h.c.\right) ,
\label{mI-2} \\
H_{fr}^{c} &=&\sum_{\sigma }\left( \omega _{0}-\omega \right) a_{0\sigma
}^{\dag }a_{0\sigma }+  \notag \\
&&\Delta \left( a_{0e}^{\dag }a_{0e}-a_{0g}^{\dag }a_{0g}\right) +2\Omega
_{I}S_{I}^{z},  \notag \\
H_{ex} &=&-2J_{z}\left( a_{0e}^{\dag }a_{0e}-a_{0g}^{\dag }a_{0g}\right)
S_{I}^{z}  \notag \\
&&-J\left( a_{0e}^{\dag }a_{0g}S_{I}^{-}+h.c.\right) .  \notag
\end{eqnarray}%
Here the superscript $\sigma \in \left\{ e,g\right\} $ in $H_{qw}^{c\sigma }$
describes the spin degree of freedom, where $e$ ($g$) denotes spin up
(down). The host lattice is modeled as tight-binding Hamiltonian, which is
represented by a sequence of potential sites with on-site energies $\omega $
except the one at $j=0$, and constant amplitude $\xi $ of hopping between
the neighboring sites. $a_{j\sigma }^{\dag }$ ($a_{j\sigma }$) is the
creation (annihilation) operator for the 1D wire. The magnetic impurity of
spin $S_{I}$ is located only at one site, which is chosen as the origin. A
given spin state of the impurity is changed by the spin rising operator $%
S_{I}^{+}$ and lowing operator $S_{I}^{-}$. In Eq.~(\ref{mI-2}), Hamiltonian
$H_{fr}^{c}$ has taken the interaction energy of the spin magnetic moments
with the magnetic field into account through the terms $\Delta \left(
a_{0e}^{\dag }a_{0e}-a_{0g}^{\dag }a_{0g}\right) $ for the propagating
electrons and $2\Omega _{I}S_{I}^{z}$ for the impurity. Here, $\Delta $ and $%
\Omega _{I}$ are the Zeeman splitting of the energy levels in the presence
of an external magnetic field along the z-direction. The exchange
interaction among spins, which is a short range point-like interaction
localized at the point $j=0$, is described by Hamiltonian $H_{ex}$. Here we
introduce the anisotropic spin-spin interaction with XXZ-type. Since we are
interested only in the coupling to single impurity, the interaction here is
written in terms of the single spin operator above, rather than the
second-quantized description of the impurity. By the Fourier transform, the
Hamiltonian for the conduction electrons reads%
\begin{equation*}
H_{qw}^{c\sigma }=\sum_{k}\left( \omega +2\xi \cos k\right) a_{k\sigma
}^{\dag }a_{k\sigma }
\end{equation*}%
where $a_{k\sigma }^{\dag }$ and $a_{k\sigma }$ are the usual creation and
annihilation operators of the conduction electron with wave vectors $k$ and
spin $\sigma $. And the anisotropic spin-spin interaction reads%
\begin{eqnarray*}
H_{ex} &=&-2\frac{J_{z}}{N}\sum_{kk^{\prime }}\left( a_{ke}^{\dag
}a_{k^{\prime }e}-a_{kg}^{\dag }a_{k^{\prime }g}\right) S_{I}^{z} \\
&&-\frac{J}{N}\sum_{kk^{\prime }}\left( a_{ke}^{\dag }a_{k^{\prime
}g}S_{I}^{-}+h.c.\right)
\end{eqnarray*}%
where $N$ is the number of sites in the lattice. $\sum_{\sigma
}H_{qw}^{c\sigma }+H_{ex}$ is the Hamiltonian for the system consisting of a
localized spin in interaction with the conduction band, which is a simpler
Kondo model when $J_{z}=J$.

\section{\label{Sec:3}controlling the propagating electron in 1D quantum wire%
}

A flying qubit is a physical realization of a qubit which moves freely and
allows information to be transported from one location to another.
Obivously, the spin of an electron propagating along the host 1D structure
acts as a flying qubit, it also interacts and exchanges information with the
magnetic impurity due to the exchange interaction. For the sake of
simplicity, we assume the magnetic impurity is a spin-1/2 particle in the
following discussion. The single-electron particle picture is employed to
study the quantum transport in a quantum wire throughout the paper. In this
section, we will first derive the eigenvalue equation, and then study the
quantum transport and discuss how to manipulate the flying qubits in this
system.
\begin{figure}[tbp]
\includegraphics[bb=86 487 514 690, width=7 cm,clip]{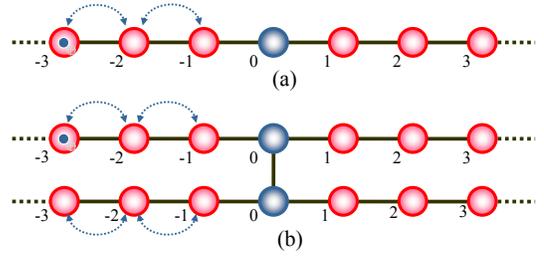}
\caption{(Color online). The reduced one-dimensional quantum wire: (a) in
the subspace with all spins up or all spins down; (b) in the subspace with
one spin up and one spin down.}
\label{scatfly:2}
\end{figure}

\subsection{spin selective scattering}

The total spin in $z$ direction is always conserved during the time
evolution of an arbitrary state. However, the applied magnetic field breaks
the spin degeneracy. Therefore, dependent on the spin-orientation of both
the Bloch electron and the magnetic impurity, the Hamiltonian in Eq.~(\ref%
{mI-1}) can be rewritten as the direct sum of three parties: $H_{ee}$, $%
H_{gg}$ and $H_{eg}$. Here,
\begin{equation}
H_{ee}=H_{qw}^{ce}+\Omega _{I}+\delta _{e}a_{0e}^{\dag }a_{0e}
\label{chain-1}
\end{equation}%
corresponds to the subspace with all spins up, where $\delta _{e}=\omega
_{0}-\omega +\Delta -J_{z}$.
\begin{equation}
H_{gg}=H_{qw}^{cg}-\Omega _{I}+\delta _{g}a_{0g}^{\dag }a_{0g}
\label{chain-2}
\end{equation}%
is the Hamiltonian of the subspace with all spins down, where $\delta
_{g}=\omega _{0}-\omega -\Delta -J_{z}$. The two-by-two matrix
\begin{equation}
H_{eg}=\sum_{\sigma }H_{qw}^{c\sigma }+\left[
\begin{array}{cc}
\bar{\omega}_{g}a_{0g}^{\dag }a_{0g}+\Omega _{I} & -J_{x}a_{0e}^{\dag }a_{0e}
\\
-J_{x}a_{0g}^{\dag }a_{0g} & \bar{\omega}_{e}a_{0e}^{\dag }a_{0e}-\Omega _{I}%
\end{array}%
\right]  \label{chain-3}
\end{equation}%
describes the interaction in the subspace with one spin down and one spin
up, where
\begin{subequations}
\label{chain-4}
\begin{align}
\bar{\omega}_{g}& =J_{z}+\omega _{0}-\omega -\Delta \text{,} \\
\bar{\omega}_{e}& =J_{z}+\omega _{0}-\omega +\Delta \text{.}
\end{align}%
In Eq.~(\ref{chain-3}), the off-diagonal elements present the interaction of
the electron spin with the x and y components of the impurity spins which
leads to spin-flip. Obviously, the original single-band split into subbands
with energy separations controlled by the local magnetic field applied to
the impurity. In Fig.~(\ref{scatfly:2}), we give a sketch of the equivalent
quantum wire in different subspaces. In subspace with all spins up or down,
it is a 1D chain with nearest ineraction, however in other subspace
corresponding to the mixture of the singlet and triplet states, it is two 1D
chains which get crossed at the point $j=0$.

\subsection{scattering in subspace with all spin down}

Now we study the transport property of the Bloch electron in this system.
Note that the configurations are similar in subspace $\left\{ a_{jg}^{\dag
}\left\vert 0g\right\rangle \right\} $ and $\left\{ a_{je}^{\dag }\left\vert
0e\right\rangle \right\} $, as well as the subspace described by Hamiltonian
$H_{eg}$ when $J_{x}=0$, so we take the scattering problem in the subspace
with all spin down as an example. When the Bloch electron and the impurity
are widely separated, we can separate the electronic and impurity-spin
degrees of freedom. Therefore the eigenstate for this subspace is in the
form
\end{subequations}
\begin{equation}
\left\vert E_{k}\right\rangle =\sum_{j}u_{k}^{1}\left( j\right) a_{jg}^{\dag
}\left\vert 0g\right\rangle \text{,}  \label{chain-5}
\end{equation}%
where $\left\vert g\right\rangle $ is the spin state of the magnetic
impurity, $u_{k}^{1}\left( j\right) $ is the probability density for finding
the Bloch electron with spin down at the $j$th site. The wave function $%
u_{k}^{1}\left( j\right) $ in this chain are obtained from the discrete Schr%
\"{o}dinger equation%
\begin{equation}
\left( E_{k}-\epsilon _{e}-\delta _{g}\delta _{j0}\right) u_{k}^{1}\left(
j\right) =\xi \left[ u_{k}^{1}\left( j+1\right) +u_{k}^{1}\left( j-1\right) %
\right]  \label{chain-6}
\end{equation}%
where
\begin{equation}
\epsilon _{e}=\omega -\Omega _{I}\text{.}  \label{chain-6a}
\end{equation}%
If we regard the hopping between different sites as the kinetic term in Eq.~(%
\ref{chain-6}), a delta-type potential is given rise to by the exchange
interaction between spins and the on-site energy at site $j=0$.

An incoming wave with energy $E_{k}$, incident from the left, results in a
reflected and transmitted wave. The wave functions in the asymptotic regions
on the left and right are given by%
\begin{equation}
u_{k}^{1}\left( j\right) =\left\{
\begin{array}{c}
e^{ikj}+r_{1}e^{-ikj}\text{ \ }j<0 \\
t_{1}e^{ikj}\text{ \ \ \ \ \ \ \ \ }j>0%
\end{array}%
\right.  \label{chain-7}
\end{equation}%
where $r_{1}$ and $t_{1}$ are some elements of the scattering matrix. Using
the connection condition at $j=0$%
\begin{subequations}
\label{chain-8}
\begin{eqnarray}
u_{k}^{1}\left( 0^{-}\right) &=&u_{k}^{1}\left( 0^{+}\right) \\
\left( E-\epsilon _{e}+\delta _{g}\right) u_{k}^{1}\left( 0\right) &=&\xi
\left[ u_{k}^{1}\left( 1\right) +u_{k}^{1}\left( -1\right) \right]
\end{eqnarray}%
the transmission amplitude can be found
\end{subequations}
\begin{equation}
t_{1}=\frac{2\xi i\sin k}{2\xi i\sin k-\delta _{g}}  \label{chain-9}
\end{equation}%
and the backscattering amplitude also can be obtained by the relation $%
t_{1}=1+r_{1}$, where the dispersion relation%
\begin{equation}
E_{k}=\epsilon _{e}+2\xi \cos k  \label{chain-10}
\end{equation}%
is used.

\begin{figure}[tbp]
\includegraphics[width=8 cm]{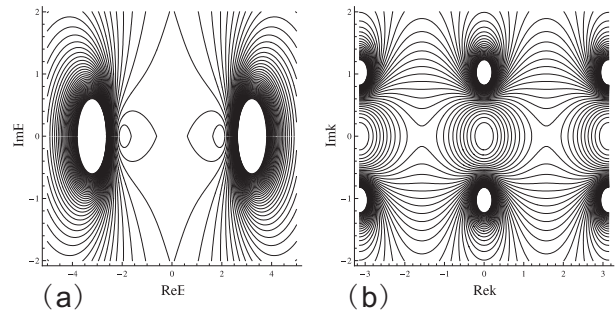}
\caption{the contour plot of the transmission coefficient in the complex
plane vs the energy $E$ in (a), the wave number $k$ in (b). $k$ is in units
of the lattice constant, $E$ is in units of hopping energy $\protect\xi $, $%
\protect\delta _{g}=\protect\sqrt{5}\protect\xi ,\protect\omega =0$. }
\label{scatfly:3}
\end{figure}

As the singularities of the scattering matrix give rise to the resonant
states, in Fig.~\ref{scatfly:3}, we plot the contour curve of the
transmission coefficient $T=\left\vert t_{1}\right\vert ^{2}$ in the complex
plane.
From Fig.~\ref{scatfly:3}(a), it can be seen that the transmission spectrum
possesses two resonance poles in the complex-energy plane. Each pole is
connected with a resonance level (or a quasibound state). According to the
Breit-Wigner formalism, the real part of a pole presents the energy of a
quasibound state, and the imaginary part is related to the lifetime of the
resonance level \cite{ArkB05,ArkB07}. In Fig.~\ref{scatfly:3}(a), these two
poles are separate on the real-energy axis, which means that each pole
corresponds to a bound state. These states are produced by lack of
periodicity in r-space. Bound states are the eigenfunctions in this system~%
\cite{scat-pole}. In this subspace with all spin down (or up), bound states
has no contribution to the quantum transport, because scattering states
survive only inside the band. However, in the subspace of antiparallel
spins, shown in Fig.~\ref{scatfly:2}(b), things become different. We will
discuss this situation in the following section. Obviously, the energy of
these bound states is tunable with gate voltage at origin applied by a
quantum point contact, and the exchange coupling along z direction as well
as the magnetic field. Transmission zeroes also include in Fig.~\ref%
{scatfly:3}(a), which appear at the border of the band. Transmission
vanishes due to the vanishing group velocities at $k=0,\pm \pi $. Figure~\ref%
{scatfly:3}(b) is the contour plot of transmittance as a function a complex
momentum. It confirms the above results. Furthermore, transmission vanishes
and two bound states appear outside the band at Re$k=0$, $\pm \pi $. This
result can be found analytically by substituting the complex momentum into
the dispersion relation in Eq.~(\ref{chain-10}). Actually the energy and the
formation of the bound states can be analytically gotten by the following
assumption%
\begin{equation}
u_{\kappa }^{1}\left( j\right) =\left\{
\begin{array}{c}
C_{1}e^{\left( in\pi +\kappa \right) j}\text{ \ }j<0 \\
C_{1}e^{\left( in\pi -\kappa \right) j}\text{ \ }j>0%
\end{array}%
\right. \text{,}  \label{chain-11}
\end{equation}%
where $n=0,1$ and $\kappa >0$. From the continuous condition given in Eq.~(%
\ref{chain-8}b), the energy of the bound state is obtained as%
\begin{equation}
E_{k}=\omega +\left( -1\right) ^{n}\sqrt{4\xi ^{2}+\delta _{g}^{2}}\text{,}
\label{chain-12}
\end{equation}%
and the normalized wavefunction reads%
\begin{equation}
u_{\kappa }^{1}\left( j\right) =\left\{
\begin{array}{c}
\sqrt{\tanh \kappa }e^{\left( in\pi +\kappa \right) j}\text{ \ }j<0 \\
\sqrt{\tanh \kappa }e^{\left( in\pi -\kappa \right) j}\text{ \ }j>0%
\end{array}%
\right. \text{.}  \label{chain-13}
\end{equation}%
Here, only bound states with even parity exist.

\subsection{spin-dependent switch}

We now investigate the scattering process of the propagating electron in a
subspace with one spin up and one spin down. The stationary state in this
excited space has the form%
\begin{equation}
\left\vert E_{k}\right\rangle =\sum_{j}\left[ u_{k}^{g}\left( j\right)
a_{jg}^{\dag }\left\vert 0e\right\rangle +u_{k}^{e}\left( j\right)
a_{je}^{\dag }\left\vert 0g\right\rangle \right]  \label{switch-1}
\end{equation}%
where $u_{k}^{i}\left( j\right) $ is the wave function of the Bloch electron
with spin $i=e$ or $g$. Applying Eq.~(\ref{switch-1}) into the stationary
Schr\"{o}dinger equation, one obtains
\begin{subequations}
\label{switch-2}
\begin{align}
J_{x}\delta _{j0}u_{k}^{e}\left( j\right) & =\xi \left[ u_{k}^{g}\left(
j+1\right) +u_{k}^{g}\left( j-1\right) \right] \\
& -\left( E_{k}-\epsilon _{g}-\bar{\omega}_{g}\delta _{j0}\right)
u_{k}^{g}\left( j\right)  \notag \\
J_{x}\delta _{j0}u_{k}^{g}\left( j\right) & =\xi \left[ u_{k}^{e}\left(
j+1\right) +u_{k}^{e}\left( j-1\right) \right] \\
& -\left( E_{k}-\epsilon _{e}-\bar{\omega}_{e}\delta _{j0}\right)
u_{k}^{e}\left( j\right)  \notag
\end{align}%
where $\bar{\omega}_{i}$ ($i=e,g$) are given in Eq.~(\ref{chain-4}), $%
\epsilon _{e}$ is listed in Eq.~(\ref{chain-6a}), the parameter
\end{subequations}
\begin{equation}
\epsilon _{g}=\omega +\Omega _{I}\text{.}  \label{switch-2b}
\end{equation}%
Equation (\ref{switch-2}) shows that there is a delta potential at $j=0$ in
each channel, which can be adjusted by on-site energy $\omega _{0}$, the
Zeeman energy $\Delta $, and the exchange interaction strength $J_{z}$
almong $z$ direction. Obviously, when $J_{x}=0$, these two chains are
independent.\ From the above discussion, we have already known that a
discrete state is created by the impurity and this discrete energy is
outside the band. In this section, new features arise due to the
nonvanishing coupling $J_{x}$. It is well known that when a system is
characterized by a coupling of a certain discrete energy and a continuum
state, Fano resonance~\cite{Fano,ArkB05,SETB00} appears because the discrete
state offers one additional propagation path in the wave scattering which
interact constructively or destructively.

\begin{figure}[tbp]
\includegraphics[width=8 cm]{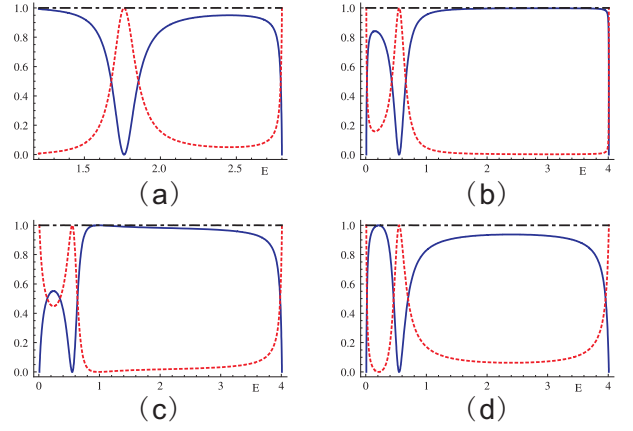}
\caption{(Color online). The transmission $T_{rl}^{g}=|t_{rl}^{g}|^{2}$
(blue solid line) and reflection coefficient $R_{ll}^{g}=|r_{ll}^{g}|^{2}$
(red dotted line), and the sum of $T_{rl}^{g}+R_{ll}^{g}$ (black dot-dashed
line) as a function of either the energy $E$. Here $\protect\omega=\protect%
\omega_{0}=0,J_{x}=0.5$. Parameters are in units of $\protect\xi $. (a) $%
\Omega _{I}=0.8,\Delta=0.8,J_{z}=0.8$, (b) $\Omega _{I}=2.01,
\Delta=0.8,J_{z}=0.8$, (c) $\Omega _{I}=2.01,\Delta=1,J_{z}=0.6$, (d) $%
\Omega _{I}=2.01,\Delta=0.6,J_{z}=1$.}
\label{scatfly:4}
\end{figure}
A propagating electron in these channel will occupy an energy of the form
\begin{equation}
E_{k}=\epsilon _{g}+2\xi \cos k_{g}=\epsilon _{e}+2\xi \cos k_{e}\text{,}
\label{switch-2a}
\end{equation}%
where $k_{e}$ and $k_{g}$ are the electron wave vectors. The process that an
incident wave impinges upon the structure under study, and transmitted and
reflected wave emerge, is formulated by
\begin{subequations}
\label{switch-3}
\begin{align}
u_{k}^{g}\left( j\right) & =\left\{
\begin{array}{c}
e^{ik_{g}j}+r_{ll}^{g}e^{-ik_{g}j}\text{ \ }j<0 \\
t_{rl}^{g}e^{ik_{g}j}\text{ \ \ \ \ \ \ \ \ }j>0%
\end{array}%
\right. \text{,} \\
u_{k}^{e}\left( j\right) & =\left\{
\begin{array}{c}
De^{-ik_{e}j}\text{ \ \ \ \ \ \ \ \ }j<0 \\
Be^{ik_{e}j}\text{ \ \ \ \ \ \ \ \ \ }j>0%
\end{array}%
\right. \text{,}
\end{align}%
Applying Eq.~(\ref{switch-3}) to the discrete Schr\"{o}dinger equation~(\ref%
{switch-2}) for the $0$th and $\pm 1$th sites, we immediately obtain the
transmission amplitude
\end{subequations}
\begin{equation}
t_{rl}^{g}=\frac{2i\xi \sin k_{g}\left( 2i\xi \sin k_{e}+\bar{\omega}%
_{e}\right) }{\left( 2i\xi \sin k_{g}+\bar{\omega}_{g}\right) \left( 2i\xi
\sin k_{e}+\bar{\omega}_{e}\right) -J_{x}^{2}}\text{,}  \label{switch-4}
\end{equation}%
within the channel, where the spins of the Bloch electron and the magnetic
impurity are down and up respectively. For later convenience, we denote this
channel as $\left\vert ge\right\rangle $ channel. The other channel in this
subspace is denoted by $\left\vert eg\right\rangle $. The transmission
amplitude from the $\left\vert ge\right\rangle $ channel to $\left\vert
eg\right\rangle $ channel reads%
\begin{equation}
B=\frac{2i\xi J_{x}\sin k_{g}}{\left( 2i\xi \sin k_{g}+\bar{\omega}%
_{g}\right) \left( 2i\xi \sin k_{e}+\bar{\omega}_{e}\right) -J_{x}^{2}}\text{%
.}  \label{switch-5}
\end{equation}%
The reflection amplitudes $r_{ll}^{g}$ within the $\left\vert
ge\right\rangle $ channel and $D$ from the $\left\vert ge\right\rangle $
channel to $\left\vert eg\right\rangle $ channel can be obtained by the
relation $t_{rl}^{g}=1+r_{ll}^{g}$ and $D=B$ respectively. Here, we interest
in a specific situation, namely, a regime where the Fano resonance might
happen. Due to the removing the degeneracy of these two channels by the
applied magnetic field, we can study the above regime by replacing $k_{e}$
with $n\pi +i\kappa _{e}$. Then Eq.~(\ref{switch-4}) becomes%
\begin{equation}
t_{rl}^{g}=\frac{2i\xi \sin k_{g}\left[ \bar{\omega}_{e}-\left( -1\right)
^{n}2\xi \sinh \kappa _{e}\right] }{\left( 2i\xi \sin k_{g}+\bar{\omega}%
_{g}\right) \left[ \bar{\omega}_{e}-\left( -1\right) ^{n}2\xi \sinh \kappa
_{e}\right] -J_{x}^{2}}  \label{switch-6}
\end{equation}%
In the following discussion we set $n=0$.

\begin{figure}[tbp]
\includegraphics[width=4 cm]{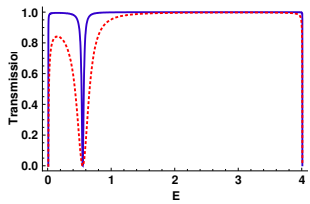}
\caption{(Color online). The transmission coefficient as a function of the
energy $E$. Here $\protect\omega=\protect\omega_{0}=0,\Omega
_{I}=2.01,\Delta=0.8,J_{z}=0.8$, $J_{x}=0.2$ for blue solid line and $%
J_{x}=0.5$ for red dotted line in units of $\protect\xi $.}
\label{scatfly:5}
\end{figure}
In Fig.~\ref{scatfly:4}, we plot the transmission coefficient $%
T_{rl}^{g}=|t_{rl}^{g}|^{2}$ (blue solid line), reflection coefficient $%
R_{ll}^{g}=|r_{ll}^{g}|^{2}$ (red dotted line), and the sum of $%
T_{rl}^{g}+R_{ll}^{g}$ (black dot-dashed line) as a function of the energy $%
E $. It can be found that: 1) there is a transmission zero (or a perfect
reflection) inside the band; 2) the flow is preserved, which is implied by $%
T_{rl}^{g}+R_{ll}^{g}=1$ in this regime. The vanishing transmission is
caused by the coupling between the continuum and the bound state. The reason
for the conserved flow is that except the bound state, other states are
forbidden outside the band of $\left\vert eg\right\rangle $ channel, which
is the energy regime we are interested, therefore energy is conserved in $%
\left\vert ge\right\rangle $ channel. The difference between Fig.~\ref%
{scatfly:4}(a) and others are that there is some overlap of the continuum
between these two channels, which is why there are only two transmission
zeroes in Fig.~\ref{scatfly:4}(a) and three transmission zeroes in other
figures. From Eq.~(\ref{switch-2}b), we know that there are no delta
potentials at the $0$th site in Fig.~\ref{scatfly:4}(a,b), however there are
a delta potential well in Fig.~\ref{scatfly:4}(c) and a delta potential
barrier in Fig.~\ref{scatfly:4}(d). The well increase the transmission at
the right side of the dip and decrease it at the left side of the dip, vice
verse for barrier. Figure~\ref{scatfly:5} shows the influence of the
coupling strength $J_{x}$. It can be found that the width of the dip
increases as $J_{x}$ increases. The above discussion implies that one can
use the local magnetic field and gate voltage to control the transport of
this flying qubit, a device with filtering and switching function is
obtained.

\section{conclusion}


We have investigated the spin of an electron propagating along the host 1D
structure, which interacts and exchanges information with a local magnetic
impurity. By the tight-binding approach, the spin-dependent
transmission/reflection coefficients is calculated within single-electron
configurations. The evanescent states are found, due to the breaking
periodicity of this 1D structure in r-space, and the energies of the
evanescent states can be adjusted by the local electrode and the external
magnetic field. In particular, we have shown that a total reflection occurs
for a propagating electron spin incident with the energy of the evanescent
state and anti-parallel to the spin orientation of the impurity. This total
reflection is originated from the Fano resonance. Therefore, the Zeeman
splitting and spin-flip interaction are necessary.

\section{Acknowledgments}

The author thanks Arkady M. Satanin for useful discussions. This work is
supported by the NSFC under Grants No.~10704023 and 10775048, the NFRPC with
under Grant Nos. 2007CB925204, New Century Excellent Talents in University
(NCET-08-0682), and Scientific Research Fund of Hunan Provincial Education
Department No.~09B063 and No.~09C638

\appendix

\end{document}